\documentclass[12pt]{article}  
\usepackage{graphicx}
\newcommand{\vv}{\vspace*{1.5ex}}   
                    
\newcommand{\hh}{\hspace*{4mm}}     
                            \newcommand{\no}{\noindent}
 \newcommand{\bc}{\begin{center}}
 \newcommand{\ec}{\end{center}}
                   \newcommand{\bfr}{\begin{flushright}}
                   \newcommand{\efr}{\end{flushright}}
   \newcommand{\ii}{\item}
     \newcommand{\be}{\begin{enumerate}}
     \newcommand{\ee}{\end{enumerate}}
        \newcommand{\bi}{\begin{itemize}}
        \newcommand{\ei}{\end{itemize}}
            \newcommand{\bd}{\begin{description}}
            \newcommand{\ed}{\end{description}}
                \newcommand{\beq}{\begin{equation}}
                \newcommand{\eeq}{\end{equation}}
                  \newcommand{\bea}{\begin{eqnarray}}
                  \newcommand{\eea}{\end{eqnarray}}

      \newcommand{\bfi}{\begin{figure}}
      \newcommand{\efi}{\end{figure}}
\newcommand{\bay}{\begin{array}{l}}
\newcommand{\eay}{\end{array}}
            \newcommand{\dd}{\mbox{d}}
\def\mb#1{\mbox {\boldmath {$#1$}}} 

    \newcommand{\pa}{\partial}
    
    \newcommand{\Del}{\Delta}

    \newcommand{\sig}{\sigma}
    \newcommand{\eps}{\epsilon}
    \newcommand{\ga}{\gamma}
    \newcommand{\Ga}{\Gamma}
    \newcommand{\om}{\omega}

\newcommand{\vbf}{\mbox{\boldmath $v$}}
        
\begin{document}   
      \thispagestyle{empty} 
      \textwidth 6.5in \oddsidemargin -.0in \textheight 9.5in \topmargin -.5in

        \hspace*{1mm}  \vspace*{-0mm}
\noindent {\footnotesize {{\em \hfill      Submitted to www.arxiv.org  \\ 
\hspace*{1mm}  \hfill June 4, 2013 }}
\mbox{ } \vskip 1.5in
\begin{center}
{\Large {\bf 
  Constitutive Model for Material \\[2.5mm] Comminuting at High Shear Rate
           }}\\[20mm]
{\large {\sc Zden\v ek P. Ba\v zant and Ferhun C. Caner}}
\\[1.5in]

{\sf Report No. 13-05/732c}\\[2.2in]

McCormick School of Engineering and Applied Science
\\ Departments of Civil and Mechanical Engineering, and Materials Science
\\ Northwestern University
\\ Evanston, Illinois 60208, USA
\\[1in]  {\bf May 28, 2013} 
\end{center}
\clearpage   \pagestyle{plain} \setcounter{page}{1}


\begin{center}
 {\Large {\sf   
      Constitutive Model for Material Comminuting at High Shear Rate}} \\[7mm]
{\large {\sf
    Zden\v ek P. Ba\v zant\footnote{
McCormick Institute Professor and W.P. Murphy Professor of Civil
Engineering and Materials Science, Northwestern University, 2145 Sheridan
Road, CEE/A135, Evanston, Illinois 60208; z-bazant@northwestern.edu.} and
    Ferhun C. Caner\footnote{
Institute of Energy Technologies (INTE/ETSEIB), Universitat Politecnica de
Catalunya, Campus Sud, 08028 Barcelona, Spain, e-mail:ferhun.caner@upc.edu,
and Visiting Scholar, Dept. of Civil and Environmental Engineering,
Northwestern University, 2145 Sheridan Rd., Evanston, IL 60208, USA}
 }}
\end{center} \vskip 5mm   

\noindent {\bf Abstract:}\, {\sf
  The modeling of high velocity impact into brittle or quasibrittle solids is hampered by the unavailability of a constitutive model capturing the effects of material comminution into very fine particles. The present objective is to develop such a model, usable in finite element programs. The comminution at very high strain rates can dissipate a large portion of the kinetic energy of an impacting missile. The spatial derivative of the energy dissipated by comminution gives a force resisting the penetration, which is superposed on the nodal forces obtained from the static constitutive model in a finite element program. The present theory is inspired partly by Grady's model for comminution due to explosion inside a hollow sphere, and partly by analogy with turbulence. In high velocity turbulent flow, the energy dissipation rate is enhanced by the formation of micro-vortices (eddies) which dissipate energy by viscous shear stress. Similarly, here it is assumed that the energy dissipation at fast deformation of a confined solid gets enhanced by the release of kinetic energy of the motion associated with a high-rate shear strain of forming particles. For simplicity, the shape of these particles in the plane of maximum shear rate is considered to be regular hexagons. The rate of release of free energy density consisting of the sum of this energy and the fracture energy of the interface between the forming particle is minimized. This yields a relation between the particle size, the shear strain rate, the fracture energy and the mass density. The particle sizes are assumed to be distributed according to Schuhmann's power law. It is concluded that the minimum particle size is inversely proportional to the (2/3)-power of the shear strain rate, that the kinetic energy release is to proportional to the (2/3)-power, and that the dynamic comminution creates an apparent material viscosity inversely proportional to the (1/3)-power of the shear strain rate. The effect of dynamic comminution can be simply taken into account by introducing this apparent viscosity into the constitutive model such as  microplane model M7.
 }

\subsection*{Introduction}

In spite of the recent advances in the constitutive modeling of concrete, the finite element models for impact of missiles onto the walls of concrete structures severely overestimate the depth of penetration and, in the case of perforation, the exit velocity. By contrast with impact on thin walls ($<10$ cm) or high rate tensile fracture, the penetration of thick walls causes comminution of a significant portion of concrete into fine particles (0.01 mm--1 mm). Inclusion of the viscoelastic rate effect and the effect of crack growth rate does not suffice by far for obtaining correct predictions. The underestimation of the exit velocities and penetration depths is severe even when the finite element code uses a highly realistic constitutive model such as the new microplane model M7 \cite{CanBazM7, CanBaz13}, which is an improvement of model M4 \cite{BazCan-00} and provides very good fits of virtually the complete range of the experimental data from diverse types of uniaxial, biaxial and triaxial tests, including the tests of vertex effect and of the compression-shear behavior under very high confinement.

The macroscopic constitutive equation with softening damage, calibrated by standard laboratory tests at low strain rates, cannot describe material comminution into sub-mesoscale particles. It can capture only the energy dissipation by  meso-scale fragmentation, i.e., the creation of fragments of the same order of magnitude as the dominant meso-scale material inhomogeneities (such as the largest aggregate pieces in concrete). This is the only kind of comminution that occurs in the standard laboratory tests of damage behavior. This limitation applies even if the constitutive equation is enhanced by the material rate effects (which include viscoelasticity and the effect of rate-dependent growth of mesoscale cracks).

The purpose of the present paper is to show how the constitutive model can take into account the energy dissipated by material comminution at very high strain rates, which can be very large. The spatial derivative of the energy dissipated by comminution represents a compressive force resisting the penetration, which has the effect of greatly increasing the finite element nodal forces obtained with a standard macroscopic constitutive model.

The literature on the analysis of impact is vast and great progress has been achieved in many directions \cite{Gra82, Gra85, Adl-Dan11, Adl-McD11, GatPij02, Ozb-Rei11, For-Gat08, CamOrt96, ShiNes-98, GaiEsp02, Doy02, KozOzb10, DesEva08, WeiEvaDes09, FerDesEva10, GraKip79, CadLab01}. As it appears, however, there apparently exists no constitutive law that would model the effect of material comminution and could be used to formulate an initial-boundary value problem underlying a finite element formulation. One important advance has been the development of the so-called "mescall" models \cite{Doy02,KozOzb10,DesEva08,WeiEvaDes09,FerDesEva10}. They describe the branching of individual, dynamically propagating, cracks but do not provide a constitutive model for a material comminuting into a vast number of tiny fragments.

Another advance among the computer simulation community was Adley et al.'s  \cite{Adl-Dan11, Adl-McD11} recent adaptation of a dynamic finite element code with microplane model M4 in which the strain-dependent strength limits (called stress-strain boundaries) of microplane model M4 were scaled sharply so as to fit the test data on missile penetration. However, after this kind of purely empirical adjustment, the constitutive model no longer fits the multitude of the standard uni-, bi-, and tri-axial laboratory tests of concrete by which the microplane model was calibrated. It also no longer fits the high-rate dynamic fracture tests of notched specimens conducted by O\v zbolt and Reihnardt  \cite{Ozb-Rei11}. These tests are in perfect agreement with the unadjusted microplane model M4, in which the increase of material strength with the strain rate is mild and smooth up to very high rates. The point to mote is that, in contrast to missile impact, the notched specimens of O\v zbolt and Reinhardt exhibit no material comminution. Adley et al.'s empirical adjustment thus loses prediction capability except for situations very similar to those for which the microplane strength limits have been adjusted.

The theory proposed here is partly inspired by analogy with turbulence. In high velocity turbulent flow, the energy dissipation rate is greatly enhanced by the formation of micro-vortices (eddies) which dissipate energy by viscous shear stress. By analogy, it is assumed that the energy dissipation at fast deformation of a confined solid gets greatly enhanced by the release of kinetic energy of high shear strain rate of forming particles. Another inspiration for the present model is Grady's model for an explosion in a hollow sphere \cite{Gra82}, in which the kinetic energy of volumetric strain is considered  as the driving force of comminution. Here we propose that it must be the kinetic energy of shear strain that drives the comminution.

The present paper presents the theory. The numerical verification and validation by large-scale finite element simulations of missile impact and by the simulation of Hopkinson bar tests os strength. These simulations use   the microplane model M7 with comminution enhancement, as presented in a paper by Caner and Ba\v zant which follows \cite{CanBazM7h13}.

\subsection*{Kinetic Energy Density of Comminuting Micro-Particles}

Consider first a simple idealized comminution process in which the material is comminuted to identical particles (Fig. \ref{f1}a,b,c,d). The particles must fill the space completely. At first, we consider the particles to be identical. The only possible repetitive regular subdivisions of the material in the plane of maximum shearing are the squares, isosceles triangles, and regular hexagons.

For graphical two-dimensional visualization, it is more instructive to begin discussion with a subdivision of the material into squares (or cubes), as depicted in Fig. \ref{f1}a, for the undeformed state. During deformation, coordinate $x$ and $y$, along with the lines of the squarer subdivision (yet to occur), rotate by angle $\om$. Simultaneously, these lines and also get skewed relative to axes $x$ and $y$ by angles $\eps_D$ representing pure shear (Fig. {f1}b).

At a certain moment, the strain rate becomes high enough for the kinetic energy of deforming material to suffice for creating the fractures that rapidly comminute the material into separate identical square particles shown in Fig. \ref{f1}c. As that happens, the particles regain their original undeformed shape, i.e., become squares again, while the centers of the comminuted particles still conform to the same macroscopic displacement field, which means that the dashed lines connecting these center are identical in Fig. \ref{f1} b and c (we ignore the necessary crushing of the particle corners since it is a second-order small correction).

As the particles return to their near-original shape, they release their kinetic energy $\Del {\cal K}$ while the opposite faces of neighboring particles slip against each other, as marked by double arrows in Fig. \ref{f1}c.

The global kinetic, energy which excludes the kinetic energy of shear strain rate of the particles and is defined as
 \beq
  \bar K = \sum_i h^3 \,\frac \rho 2\; (\dot u_0^2 + \dot v_0^2)_i
 \eeq
(where $\dot u_0, \dot v_0$ are the velocity components of the centers $i=1,2,3,...$ of the particles or of the eddies, and $h$ = side of the squares) remains unchanged as the particles separate.

A more correct choice is a subdivision into the hexagons (Fig. \ref{f2}a) because it gives, in two dimensions, the smallest surface-volume ratio. In the third dimension we assume the particles to be hexagonal prisms of a length equal to the corner-to-corner diameter $h$ of the hexagon. The volume of one particle and the interface area $S$ of all the particles per unit volume of material are, respectively,
 \bea \label{Vp}
  V_p &=& c_v h^3,~~~c_v = \frac{3 \sqrt{3}} 8 \doteq 0.6495
 \\ \label{S}
  S &=& \frac{c_s} h,~~~~~
  c_s = \frac 3 8 \left( 2 + \sqrt{3} \right) \doteq 2.155
 \eea
where $c_s$ = dimensionless constant. Note that since each two neighboring particles share the same interface, $S = \frac 1 2 S_p\, / V_p$ where $S_p$ = surface area of one hexagonal prism.

Let the Cartesian coordinates $x, y, z$ be placed so that plane $(x,y)$ be the the plane of maximum shear strain rate among all possible orientations.
The field of displacements is considered to consist of pure shear strain $\eps_D$ and rotation $\om$. The maximum shear strain rate, denoted as $\dot \eps_D$, is chosen to represent pure shear, in which case $\dot \eps_D$ also represents the effective deviatoric strain rate, i.e.,
 \beq \label{epsD}
   \dot \eps_D = \sqrt{\dot {\eps_D}_{ij} \dot {\eps_D}_{ij}}
 \eeq
while the shear angle rate is $\dot \ga = 2 \dot \eps_D$ (note that repetition of tensorial subscripts implies summation). Here ${\eps_D}_{ij}$ is the deviatoric strain tensor in Cartesian coordinates, and the superior dots denote the time derivatives.

At first the hexagonal cells, not yet separated (Fig. \ref{f2}b), undergo shear strain $\eps_D$ in conformity with the material as a whole. However, a quasibrittle material such as concrete cannot undergo a large strain and so, at a certain moment, the material will fracture into the hexagonal prisms. This sudden fracturing releases the strain in the hexagonal particles and allow them to regain their original shape. As they do, they must rotate against each other by angle $\eps_D$ (Fig. \ref{f2}c).

It may be assumed that, during this whole process, the particle centers move so as to follow the field of macroscopic pure shear strain $\eps_D$ and simultaneous material rotation rate $\dot \om$ (Fig. \ref{f2}c) in plane $(x,y)$. The effect of hydrostatic pressure will be included later.
Explosive volume expansion, which leads to a different type of comminution, is not considered.

Before comminution, the displacement velocities $\dot u$ and $\dot v$ in the directions of current (Eulerian) coordinates $x$ and $y$ whose origin is placed into the particle center are
 \beq \label{uv}
  \dot u = \dot u_0 - \dot \om y + \dot \eps_D y,~~~
  \dot v = \dot v_0 + \dot \om x + \dot \eps_D x
 \eeq
After fracturing and separation of the particles these velocities change to
 \beq \label{uv+}
  \dot u^+ = \dot u_0 - \dot \om y,~~~
  \dot v^+ = \dot v_0 + \dot \om x
 \eeq
The drop in kinetic energy of each hexagonal cell is
 \bea   \label{DelVK}
  - V_p \Del{\cal K} &=& h \int_A \frac \rho 2 \left( \dot u^2 + \dot v^2 -
  (\dot u^+)^2 - (\dot v^+)^2 \right)\, \dd A
 \\  \label{DelVK'}
  &=& \frac{h \rho} 2 \int_A (x^2 + y^2)\ \dd A\ \dot \eps_D^2
  = \frac{h \rho} 2 \int_A r^2\ \dd A\ \dot \eps_D^2
  = \frac{h \rho} 2\ I_p\, \dot \eps_D^2
  \eea
or, per unit volume of material,
 \beq \label{DelK}
  \Del {\cal K} = - c_k \rho h^2 \dot \eps_D^2
 \eeq
where
 \beq
  I_p = \frac{5 \sqrt{3}}{128}\ h^4\, \doteq 0.06766\, h^4,~~~~
  c_k = \frac{I_p}{2 h V_p} \doteq 0.006510
 \eeq
Here ${\cal K}$ = drop of kinetic energy of the particle per unit volume, $c_k$ = dimensionless coefficient of kinetic energy, $\rho$ = mass density, $A$ = area of the hexagon, $r$ = radial coordinate, $I_p = I_x + I_y$ = centroidal polar moment of inertia of the hexagon (and $I_x, I_y$ = moments of inertia about axes $x$ and $y$, the orientation of which does not matter).

Note that, in calculating the integral in Eq. (\ref{DelVK}), the integrand terms with $x$, $y$ vanished, because of symmetry. Also note that the material rotation velocity $\dot \om$ has no effect on ${\cal K}$. This must have been expected, since rigid body rotations cannot cause fracturing.

\subsection*{Generalization to Randomly Distributed Particle Sizes}

Limiting consideration to micro-particles of one size would be an
oversimplification. It is well-known that, in all sorts of dynamic
comminution, the particle sizes vary randomly and are approximately distributed according to a power law, called the Schuhmann law \cite{Sch40, Cha57, Ouc05, Cun87} (this law was found important, e.g., for calculating the correct resisting force due to comminution of concrete slabs during the collapse of the World Trade Center towers \cite{BazLe-08}). The Schuhmann law is described by the cumulative distribution function of variable particle size $s$:
 \beq  \label{Schuh}
  F(s) = \frac{s^k - h^k}{H^k - h^k}~~~[s \in (h, H),~~F(s) \in (0,1)]
 \eeq
where $k$ = empirical constant (typically $k \approx 0.5$), $h = s_{min}$ =
minimum micro-particle size, and $H = s_{max}$ = maximum micro-particle size (usually $H/h$ = 10 to 100). This means that, within the size interval $(s, s + \dd s)$, the number of micro-particles per unit volume is $\dd F(s) / s^3$.
Hence, since for size $s$ the particle interface area per unit volume is $c_s /s$, the combined interface area of all the micro-particles or random sizes per unit volume is
 \bea \label{S}
  S &=& \int_{s=h}^H \frac{c_s} s\ \dd F(s) =\ \frac{C_s} h
 \\ \mbox{where}~~~
  C_s\ &=& \ \frac{c_s k}{k-1}\; \frac{(H/h)^{k-1} - 1}{(H/h)^k - 1}
 \eea
$C_s$ is dimensionless constant. It may be checked (with the aid of L'Hospital rule) that the limit of Eq. (\ref{S}) for $H/h \to 1$ is $c_s$, as it must.

Note that, in this calculation, we did not address the fact that particles of unequal sizes cannot all be hexagonal prisms. This means that their closest packing would leave some empty interparticle spaces and would thus require a certain volume dilation, and also that particles of diverse shapes will be created. For simplicity, we do not analyze this dilation and shape diversity.

The calculations of kinetic energy loss must also be generalized to randomly distributed particle sizes. The loss of kinetic energy of the particles of all the sizes per unit volume (dimension J/m$^3$ or N/m$^2$) may be calculated, according to Eq. (\ref{S}), as follows:
 \beq \label{KhH}
  {\Del \cal K} = -\int_{s=h}^H c_k \rho s^2 \dot \eps_D^2\ \dd F(s)
 \eeq
The integration yields:
 \bea  \label{K}
  {\Del \cal K}\ &=&\ - C_k \rho h^2 \dot \eps_D^2
 \\ \mbox{where}~~~~
  C_k\ &=&\ \frac{c_k k }{k+2}\; \frac{(H/h)^{k+2} - 1}{(H/h)^k - 1}
 \eea
$C_s$ is a dimensionless constant. For the limit case of particles of one size $h$, one may check that $\lim_{H/h \to 1} C_k = c_k$.

\subsection*{Balance of Rates of Kinetic Energy and Surface Energy}

The total energy of the comminuted particles per unit volume is
 \beq  \label{F}
  {\cal F} = {\cal K} + S \Ga
 \eeq
where $\Ga$ is the interface fracture energy of the comminuting particles (dimension J/m$^2$ or N/m). The surface of the particles
should be such that
 \beq \label{Emin}
  {\cal F} = \mbox{min}~~~\mbox{or}~~~\frac{\pa {\cal F}}{\pa S} = 0
 \eeq
which requires that
 \beq \label{dKdS}
  - \frac{\pa (\Del {\cal K})}{\pa S}
  =  - \frac{\pa (\Del {\cal K}) / \pa h}{\dd S / \dd h} = \Ga
 \eeq
This equation is similar to that used in D.E. Grady's \cite{Gra82} analysis of tensile comminution due to an explosion within the center of a hollow sphere. It is seen to be identical to the energy release criterion of
fracture mechanics. In fact, it could have been derived directly from the
condition of criticality: When comminution gets under way, the energy release
rate at a given kinematic constraint (i.e., at a given $\dot \eps_D$) must be
equal to the surface energy $\Ga$.

Alternatively, one might think of imposing an overall energy balance condition ${\cal K} = - S \Ga$. However, like in fracture mechanics in general, this would not guarantee the interface fracture to begin. Eq. (\ref{dKdS}), on the other hand, means that as soon as energy balance is satisfied incrementally, the fracture begins and cannot be stopped.

Substitution of Eqs. (\ref{K}) and (\ref{S}) into (\ref{dKdS}) and
differentiation furnishes the minimum particle size:
 \bea  \label{h}
   h &=& s_{min}
   = \left( \frac{ C_a \Ga }{ \rho\, \dot \eps_D^2 } \right)^{1/3}
 \\ \label{Ca}  \mbox{where}~~~~
  C_a &=& \frac{C_s}{2C_k}
  = \frac{c_k (k-1)}{2 c_s (k+2)}\; \frac{(H/h)^{k+2} - 1}{(H/h)^{k-1} - 1}
 \eea
Here, the maximum particle size $H$ is approximately known. It cannot be larger that the maximum size $d_a$ of the inhomogeneities (or the maximum aggregate size in concrete). In fact, it must be about one order of magnitude smaller than $d_a$, i.e.,
 \beq \label{H}
  H \; \approx \; 0.1 d_a
 \eeq
because fragmentation into particles of the same order of magnitude as $d_a$ is covered by the constitutive laws based on standard static material tests. Therefore, $H$ must be about 2 mm -- 4 mm for normal concrete, and 0.05 to 2 mm for high strength concrete. Thus, to calculate $h$, one must substitute Eq. (\ref{h}) into (\ref{Ca}) and then solve the resulting nonlinear algebraic equation for $h$ iteratively, e.g., by Newton method. The ratio $H/h$ the follows.

Substitution of Eq. (\ref{h}) into Eq. (\ref{K}) further yields:
 \beq  \label{DelKin}    
   - \Del {\cal K}\, =\, (C_0 \Ga^2 \rho)^{1/3}\ \dot \eps_D^{2/3}
  ~~~\mbox{where}~~~C_0 = C_k C_s^2 /4
 \eeq

It may also be noted that, when the high-rate shearing occurs under high hydrostatic pressure $p$, the effective value of interface fracture energy, $\Ga$, may increase as a function of $p$.

\subsection*{Is the Strain Energy Release at High Strain Rate Important?}

Strictly speaking, the strain energy density ${\cal U}$ should be added on the right-hand side of Eq. (\ref{F}), i.e.,
 \beq
  {\cal F} = {\cal U} + {\cal K} + S \Ga
 \eeq
But is that important? To check it, consider a high strain rate, $\dot \eps_D = 10^4$/s and a concrete of compressive strength 5000 psi or about 35 MPa. Also consider that, in absence of significant confining pressure, the shear strain cannot exceed the tensile strength, $\tau_0 \approx 3 MPa$, which means that the shear strain can at most be $\eps = \tau_0 /2 G$ where $G \approx$ = 1.18 GPa = shear modulus of concrete. Also consider that $h =  0.5$ mm and $H/h$ = 10. Simple calculation then shows that
 \beq  \label{KU}
  \frac{\pa {\cal U} / \pa S}{\pa {\cal K} / \pa S}\; \approx\; 0.011
 \eeq
So, indeed, the strain energy is unimportant. And this result is obtained under the assumption that the shear stress transmitted between particles drops to zero, which is certainly excessive. There is interparticle friction, and possibly the shear stress drops by less than 10\%. This would increase the ratio in Eq. (\ref{KU}) by two orders of magnitude.

Under high confining hydrostatic pressure $p$, say $p = 10 \tau_0$, the estimate in Eq. (\ref{KU}) would drop by about two orders of magnitude, but this is irrelevant, because under such confinement the shear stress transmitted between particles would probably not drop at all, due to high friction.

For low loading rates, such as $\dot \eps_D < 1$/s, the strain energy ${\cal U}$ is, of course, found to dominate. But at such rates the material does not undergo comminution.

Another point to note is that the release of strain energy ${\cal U}$ does not lead to dependence of fracturing on the strain rate. Rather, it leads to a dependence on the magnitude of applied stress or strain, which is already accounted for by the softening part of the constitutive relation.

\subsection*{Some Simple Numerical Estimates}

The precise value of interface fracture energy $\Ga$ is a difficult question. On the macroscale, a typical value of fracture energy of concrete is $G_F$ = 100 J/m$^2$ = kg / s$^2$. But it has long been known the $G_F$ of all materials is larger than the surface energy $\ga$ of the solid by several orders of magnitude. The reason is that the fracture surface is nor perfectly smooth and many energy dissipating microcracks and frictional slips inevitable occur on the size of the final fracture path. Thus, the narrower the fracture process zone, the smaller should be the ratio $\Ga /G_F$. On the other hand, the value of $G_F$ or $\Ga$ increases with the loading rate, and also can greatly increase as a result of confining pressure. Furthermore, the macroscale fracture in normal concrete tends to run around the aggregates through the weak zones of load density hardened cement paste, while in comminution, fractures must run also through the strong zones of cement paste and through the aggregate pieces. So, the precise value of the ratio $\Ga / G_F$, as well as its likely dependence on particle size $s$, would have to be obtained by calibrating the fits of dynamic comminution experiments.

For a crude estimate assume $\Ga$ to be independent of particle size and assume that $\Ga \approx 100 G_F \approx 100$ J/m$^2$. Let us also assume that the maximum size of comminuted particles is $H = 3.255$ mm, which is a reasonable value for concrete. Also consider that $\rho$ = 2300 kg/m$^3$.

Further consider the strain rate $\dot \eps_D = 5 \times 10^4$ /s, which a value seen in finite element simulations of the penetration of missiles into concrete walls. The solution of Eqs. (\ref{h}) and (\ref{Ca}) then yields the minimum particle size
 \beq
  h = s_{min} = \mbox{ 0.2398 mm},
 \eeq
and $H/h = 10$. 

Evaluating the kinetic energy per unit volume from Eq. (\ref{DelK}) for the strain rate $\dot \eps_D = 10^4$/s, we obtain $\Del {\cal K} = - 59.55 \cdot 10^6$ J/m$^3$. At the same time (since 1 J/m$^3$ = 1 MPa), this energy density represents the additional pressure due to comminution:
 \beq
  \sig = - \mbox{ 59.55 MPa (8638 psi) }
 \eeq
This pressure, which resists the penetrating missile, is {\em  in addition} to the stress obtained from the macroscopic constitutive model, with its rate effect.

\subsection*{Implementation of Kinetic Rate Effect}

Energy conservation requires that ${\cal K} = {\cal D}$ = energy dissipated
per unit volume (dimension J/m$^3$).  As the same time, since the stress
(dimension N/m$^2$) is the energy per unit volume, ${\cal K} = \sig^A$ =
additional stress due to strain rate. There are two possible approaches to
enforce this energy dissipation in a finite element program:
\\ \hh 1) Either as a body force
 \beq
  {\mb f} = \pa {\cal K}\, / \, \pa {\mb x}~~~~~({\cal K} = {\cal D} = \sig^A)
 \eeq
which gets translated into equivalent nodal forces (${\mb x}$ = global
coordinate vector formed from Cartesian coordinates $x_i, i = 1,2,3)$;
\\ \hh 2) Or as an additional stress, $\sig^A = {\cal D}$, to be implemented
in the constitutive equation. We favor this approach as it seems simpler
for programming, and also is more versatile as it allows generalization to
different types of comminution.


Two alternatives of implementing approach 2 are possible.

\subsubsection*{Alternative I. Kinetic (Apparent) Viscosity:}

Since one may write $\sig^A = {\cal D} = \eta \dot \eps$, the additional
(apparent) viscous stress $\sig^A$ may be implemented as kinetic (or
apparent) viscosity $\eta = {\cal D} / \dot \eps = {- \Del \cal K} / \dot \eps$. But this equation would be acceptable only in a uniaxial model. In a
triaxial constitutive model, the additional viscous stress should be
applied only to the stress components that represent shear, i.e., as
additional deviatoric stress components $s_{ij}^A$ (which may be imagined
to act in parallel coupling with the stresses obtained from the standard constitutive model).

To ensure tensorial invariance, powers of the tensorial components are inadmissible. Only a tensorial invariant can be raised to a power. So we use again the effective deviatoric strain rate given by Eq. (\ref{epsD}),
which equals $\sqrt{2}$ times the second invariant of the deviatoric strain rate tensor $\dot {\eps_D}_{ij}$ . Since the energy density has the same dimension as the stress, Eq. (\ref{DelKin}) may now be generalized to the deviatoric stress-strain relation:
 \beq \label{sDA}
  s_{ij}^A\, =\, \eta_D\, \dot {\eps_D}_{ij},~~~~
  \eta_D =  (C_0 \Ga^2 \rho)^{1/3}\ {\dot \eps_D}^{~-1/3}
 \eeq
in which the conditions of tensorial invariance are adhered to;  $\eta_D$ is a deviatoric kinetic (or apparent) viscosity. As a check on correctness of this tensorial form, note that substitution of $s_{ij}^A = \eta_D\, \dot e_{ij}$ into the energy dissipation expression $-\Del {\cal K} = \sqrt{s_{ij}^A s_{ij}^A}$ yields the correct viscous energy dissipation $-\Del {\cal K} = \eta \dot \eps_D$.

\subsubsection*{Alternative II. Increased Strength or Yield Limit}

Since the energy per unit volume (dimension J/m$^3$, with J = Nm) has the
same dimension as the stress (dimension N/m$^2$), a increase of strain rate
may alternatively be considered to cause an increase in the strength or yield limit in the constitutive law. In the microplane model, the compressive deviatoric boundary curve for deviatoric stress $\sig_D$, and the normal boundary for tensile normal stress $\sig_N$, respectively, should thus be scaled up by the kinetic factors:
 \beq
  r_D = (C_D \Ga^2 \rho)^{1/3} \langle \dot \eps_D \rangle^{2/3},~~~~
  r_V =  (C_V \Ga^2 \rho)^{1/3} \langle \dot \eps_N \rangle^{2/3}
 \eeq
where $\eps_D$ is now the microplane deviatoric strain component, and
$\eps_N$ the microplane normal strain component.

Modeling of the enhanced resistance to high-velocity missile penetration by the rasing of boundaries of the microplane model was attempted by Adley et al. \cite{Adl-Dan11, Adl-McD11}. They raised the boundaries in a way that allowed them to fit missile penetration data. However, in the microplane model there are many boundaries and each has a different and complicated shape. Different boundaries and and different parts of each boundary should then be raised by different ratios, reshaping the boundaries. It appears to be impossible to do that without loosing the founding of the microplane model in the static triaxial test data.

\subsection*{Generalization to Kinetic Energy of Volume Expansion}

In some situations of impact and penetration, it may be possible that rate of volumetric strain $\eps_V = \eps_{kk}/3$ is so high that its kinetic energy is significantly contributing to the comminution (such a case was shown by Grady \cite{Gra82, Gra85} in his analysis of an explosion within a hollow sphere). In that case, Eqs. (\ref{uv}) need to be generalized as
 \beq \label{uv'}
  \dot u = \dot u_0 - \dot \om y + \dot \eps_D y + \dot \eps_{Ex} x,~~~
  \dot v = \dot v_0 + \dot \om x + \dot \eps_D x + \dot \eps_{Ex} y
 \eeq
in which
 \beq \label{epsEx}
  \dot \eps_{Ex}\; =\;
  \Bigl\langle \frac \dd {\dd t}\; \Bigl\langle \frac{\eps_{kk}} 3
    \Bigr\rangle \Bigr\rangle
  =  \left\{  \begin{array}{ll}
      \dot \eps_V & \mbox{if}~\dot \eps_V > 0~\mbox{and}~\eps_V > 0 \\
      0 & \mbox{otherwise}
\end{array}   \right.
 \eeq
and $t$ = time; $\dot \eps_{Ex}$ represents the expansive strain rate, which is such that no comminution results if either the volumetric strain rate is compressive or if the volumetric strain is compressive (i.e., negative). Noting that Eq. (\ref{uv+}) remains unchanged, one finds that Eqs. (\ref{K}) and (\ref{DelKin}) for the drop of kinetic energy per unit volume of material must now be generalized as follows:
 \bea \label{K'}
  &&- {\Del \cal K}\ =\ C_k \rho h^2 (\dot \eps_D^2 + \dot \eps_{Ex}^2)
 \\ \label{DelKin'}
  &&- \Del {\cal K}\ =\
      (C_0 \Ga^2 \rho)^{1/3}\ (\dot \eps_D^2 + \dot \eps_{Ex}^2)^{1/3}
 \eea
This means that, in the further resulting equations, $\dot \eps_D^2$ must now be replaced by $\dot \eps_D^2 + \dot \eps_{Ex}^2$. Hence, Eq. (\ref{h}) must be generalized as
 \beq  \label{h'}
   h = s_{min} = \left( \frac{ C_a \Ga }{ \rho\,
   (\dot \eps_D^2 + \dot \eps_{Ex}^2)} \right)^{1/3}
 \eeq

Eq. (\ref{sDA}) must now be combined with the additional apparent tensile volumetric viscous stress, as follows:
 \beq \label{sVA}
   s_{ij}^A\, =\, \eta_D\, \dot e_{ij},~~~~\sig_V^A  = \eta_V \dot \eps_{Ex}
 \eeq
in which
 \beq
  \eta_V = (C_0 \Ga^2 \rho)^{1/3}\ \dot \eps_{Ex}^{-1/3}
 \eeq
Here $\eta_V$ is the volumetric kinetic (or apparent) viscosity.


\subsection*{Discussion}

Stress increase proportional to $\dot \eps^{2/3}$ gives an enormous rate
effect, but only at very high strain rates. To assess it, consider that the
strength or yield limit is scaled up by a rate-dependent factor, $r$, and
note that, according to the test data in the literature, the rate effect
beyond that explained by viscoelasticty and crack growth rate is detectable
only for rates $\dot \eps > 0.1$/s. Knowing this fact suffices to estimate
the rate effect magnitude.

Assuming that the error of the data is not below 1\%, we have $0.1^{2/3} r
= 0.01$, from which $r = 0.0464$. So, e.g., for strain rate $10^4$/s one
has $r = (10^4)^{2/3} \times 0.0464 = 21.5$. This is the ratio by which the
stress must be increased, either by means of kinetic viscosity or by
scaling up the strength or yield limits.

The foregoing analysis of kinetic energy balance indicates that the
material must be comminuted to particles of a certain size as soon as a
certain strain rate is imposed. Partially formed particle are not
considered since the particle size $a$ is calculated from the surface area
$S$ that follows from the kinetic energy ${\cal K}$. Whether this is an
acceptable simplification is not clear. If the surfaces of particles were
only partially formed, the particle size $h$ predicted from a given $S$
would be smaller and the kinetic energy of their straining would be less.

When the shear strain rate $\dot \eps_D$ decreases, the rate effect drops,
although the particle surfaces do not disappear. When the strain rate is
raised again to the same value as before, the same strengthening due to
rate effect is predicted by the foregoing equations, even though no
additional particle surface $S$ needs to be formed and the material is
already reduced to a sort of fine sand. There seem to be no test data on
this prediction. However, if the shearing occurs while this sand is under
volumetric compression, there must be frictional slip between the
particles. The drop of slip resistance from static to dynamic friction is a
similar phenomenon as fracture and requires an energy input, which would
lead to a similar analysis as above.

\subsection*{Conclusions}
 \be  \setlength{\itemsep}{-1.4mm} \ii
The kinetic energy of motion associated with high shear strain rates ($>10^2$/s) is sufficient to provide the surface fracture energy necessary for comminution of materials such as concrete into fine particle.
 \ii
Vice versa, the fracture during comminution dissipates energy that suffices for significant deceleration of a missile penetrating the material.
 \ii
At very high strain rates, kinetic energy of strain rate is orders of magnitude higher than the strain energy.  Hence, the classical fracture mechanics does not apply.
 \ii
At high shear rates, the release of strain energy due to fracturing is far smaller than the release of kinetic energy (at strain rate $10^4$/s it is about 100-times smaller, and thus negligible).
 \ii
The comminuted particles lose their shear strain and subsequent shearing of the comminuted material occurs by frictional interparticle slip, like in sand.
 \ii
To calculate the particle sizes, one may assume Schuhmann power law distribution of particle size. The minimum particle size then follows from the condition that the rate of release of kinetic energy of shear strain must be equal to the rate of energy dissipation due to growing area of interparticle fractures. The minimum particle size predicted for missile impact at 300 m/s is of the order of 0.1 mm.
 \ii
The present theory predicts that the density of kinetic energy available for comminution is proportional to the $(2/3)$ power of the shear strain rate, the minimum particle size is inversely proportional to the $(2/3)$ power of that rate, and the energy dissipation by comminution is equivalent to a shear viscosity decreasing as the $(-1/3)$ power of that rate. For a strain rate increase from 1/s to $10^4$/s, the result is a roughly 20-fold increase of apparent material strength due to comminution.
 \ii
Although not too important for missile impact, the theory can be extended in an analogous way by including particle comminution due to kinetic energy of volumetric strain rate.
 \ii
The theory leads to a rate-dependent modification of a constitutive equation such as the microplane model, which is easily implemented in an explicit finite element program.
 \ee

\vv \no {\small {\bf Acknowledgment:}\,  {\sf Financial support under grant
W911NF-09-1-0043/P00003 from the U.S. Army Research Office, Durham, to
Northwestern University is gratefully acknowledged. So is an additional
support for theoretical studies granted to Northwestern University by the
Agency for Defense Development (ADD), Korea.} }

\subsection*{Appendix: Partial Analogy with Turbulence}

It is interesting that the kinetic energy ${\cal K}_{shear}$ of a particle deforming by pure shear at rate $\dot \eps_D$ happens to be the same as the kinetic energy ${\cal K}_{eddy}$ of the same particle rotating as a rigid body at angular rate $\dot \om =  \dot \eps_D$. Indeed, the squares of the velocity magnitudes in the comminuting particle with shear strain rate $\dot \om_D$ and in a turbulence vortex (eddy) treated approximately as a rigid domain with angular flow velocity $\dot \om$ are, respectively,
 \bea \label{Kshear}
  {\cal K}_{shear} &=& \int_A \frac \rho 2\, |\vbf|^2 \dd A
     = \int_A \frac \rho 2\, [( \dot \eps_D y )^2 + (\dot \eps_D x)^2] \dd A
 \\ \label{Keddy}
  {\cal K}_{eddy} &=& \int_A \frac \rho 2\, |\vbf|^2 \dd A
     = \int_A \frac \rho 2\, [( \dot \om y )^2 + (\dot \om x)^2] \dd A
 \eea
Now note that when $\dot \om =  \dot \eps_D$,
 \beq
  {\cal K}_{shear} = {\cal K}_{eddy}
 \eeq
even though the velocity vectors at the corresponding points do not have the same directions.

This observation reveals a partial analogy with turbulence \cite{TenLum72}, which is what inspired the present theory. In both comminution and turbulence, the micro-level kinetic energy (Eq. (\ref{Kshear}) or (\ref{Keddy})) augments the kinetic energy of the macro-level part of the turbulent flow of a fluid, or the macrolevel kinetic energy of the assembly of the comminuting particles, which in both cases is equal to
 \beq
  \sum_i  \frac \rho 2\, (\dot u_0^2 + \dot v_0^2)_i
 \eeq
where $\dot u_0, \dot v_0$ denote the velocity components of the centers $i=1,2,3,...$ of the particles or of the eddies. The micro-level kinetic energy is dissipated by fluid viscosity in the eddies of turbulent flow, or the by the energy of interface fracture of the comminuting particles. In both cases, minimization of the total energy of motion requires requires a micro-level energy dissipation mechanism, eddy formation or comminution.

\listoffigures  
 \clearpage

 \bfi 
 \centering
 \includegraphics[width=\textwidth]{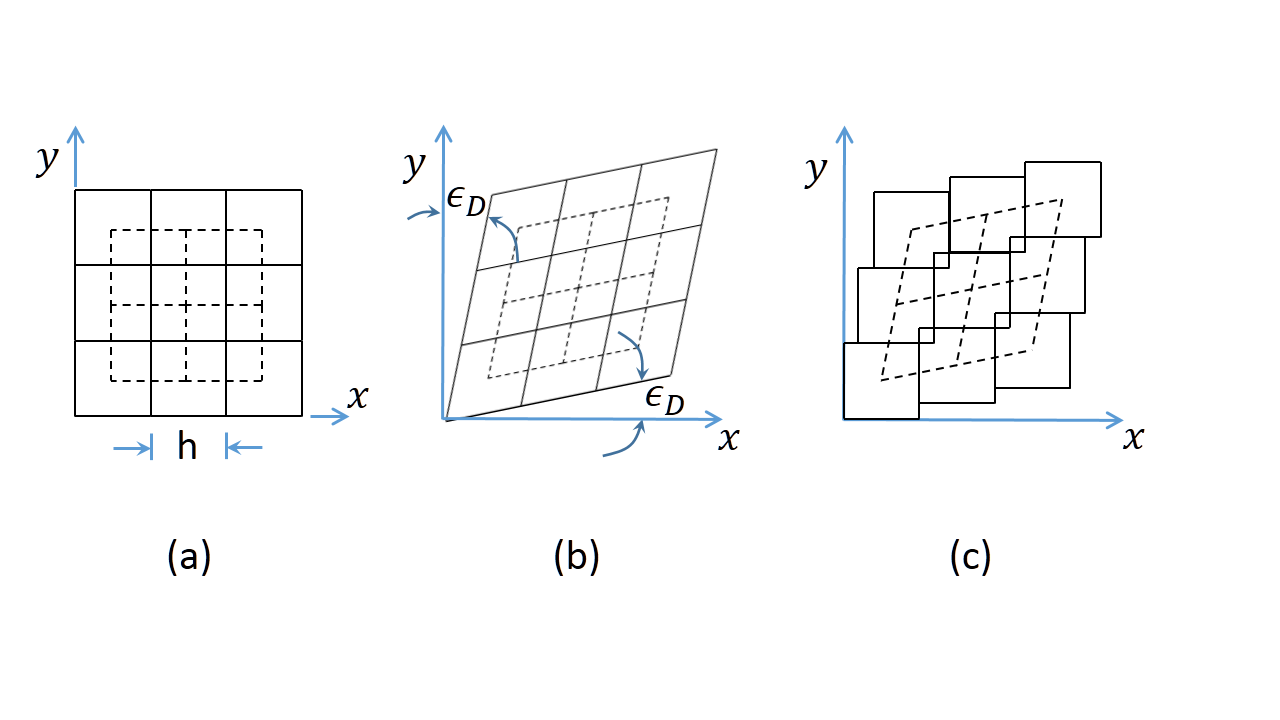}
 \caption{ \label{f1} Comminution of material into prismatic square
   particles, showing the velocities in terms of displacements during
   time $\Delta t$ (note that displacements are supposed to be
   infinitesimal, in which case the overlaps at the square corners are
   second-order small and thus negligible).}  

\efi

 \bfi 

 \centering
 \includegraphics[width=\textwidth]{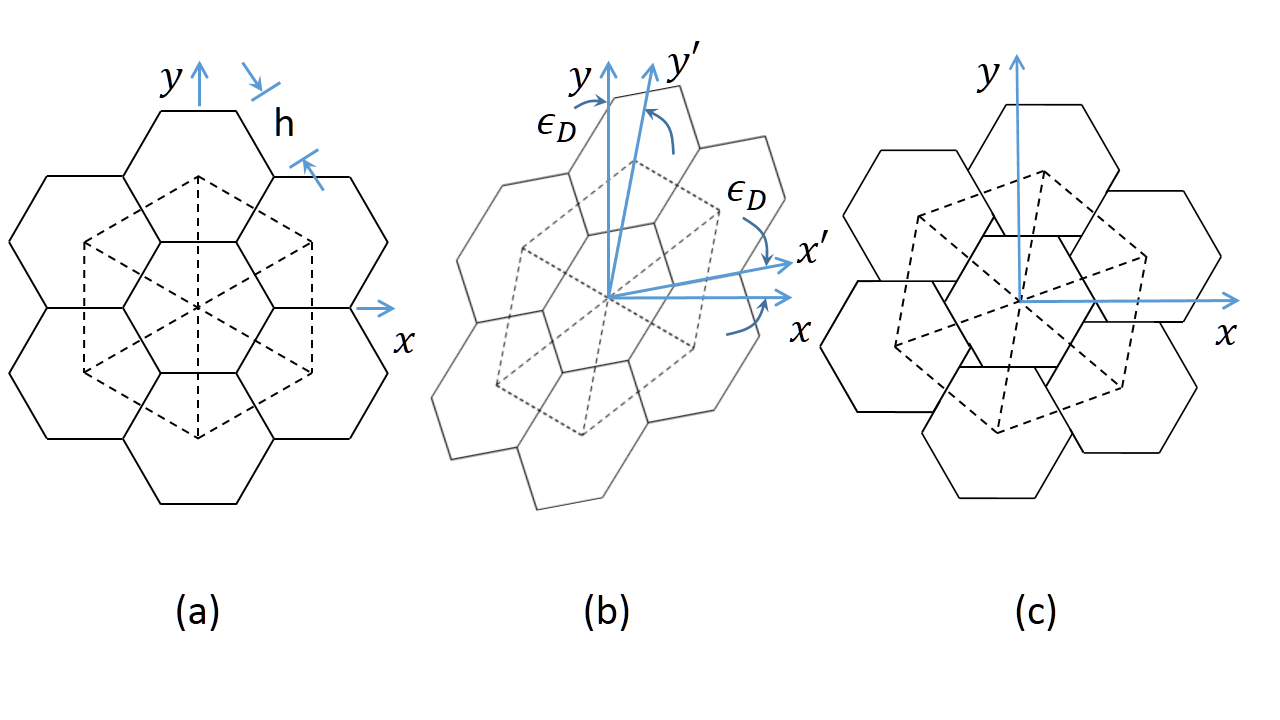}
 \caption{ \label{f2} Comminution of material into prismatic hexagonal
   particles, showing the velocities in terms of displacements during
   time $\Delta t$ (note that displacements are supposed to be
   infinitesimal, in which case the gaps at the hexagon corners are
   second-order small and thus negligible).  }  

\efi

\end{document}